\documentclass[conference]{IEEEtran}

\usepackage{algorithm}
\usepackage{algpseudocode}

\usepackage{graphicx}
\graphicspath{ {./} }
\usepackage[english]{babel}
\usepackage{blindtext}
\usepackage[inline]{enumitem}
\usepackage{enumitem}
\usepackage{balance}
\usepackage{cite}
\usepackage{amsmath,amssymb,amsfonts}
\usepackage{subcaption}
\usepackage[font=footnotesize]{caption}
\usepackage{textcomp}
\usepackage{xcolor}

\def\BibTeX{{\rm B\kern-.05em{\sc i\kern-.025em b}\kern-.08em
		T\kern-.1667em\lower.7ex\hbox{E}\kern-.125emX}}

\newcommand{\her}{\mathsf{H}}

\usepackage{setspace}

\begin{document}
	
\title{Digital Twin Aided Channel Estimation: Zone-Specific Subspace Prediction and Calibration}
\author{\IEEEauthorblockN{Sadjad Alikhani and Ahmed Alkhateeb}
	\IEEEauthorblockA{Wireless Intelligence Lab, Arizona State University, USA}{Emails: \{alikhani, alkhateeb\}@asu.edu}}
\maketitle

\begin{abstract}
Effective channel estimation in sparse and high-dimensional environments is essential for next-generation wireless systems, particularly in large-scale MIMO deployments. This paper introduces a novel framework that leverages digital twins (DTs) as priors to enable efficient zone-specific subspace-based channel estimation (CE). Subspace-based CE significantly reduces feedback overhead by focusing on the dominant channel components, exploiting sparsity in the angular domain while preserving estimation accuracy. While DT channels may exhibit inaccuracies, their coarse-grained subspaces provide a powerful starting point, reducing the search space and accelerating convergence. The framework employs a two-step clustering process on the Grassmann manifold, combined with reinforcement learning (RL), to iteratively calibrate subspaces and align them with real-world counterparts. Simulations show that digital twins not only enable near-optimal performance but also enhance the accuracy of subspace calibration through RL, highlighting their potential as a step towards \textit{learnable digital twins}.

\end{abstract}

\section{Introduction}

Efficient channel estimation is crucial for multi-antenna wireless communication systems, particularly in sparse environments where limited scatterers and dominant line-of-sight components characterize the channel \cite{7400949}. This sparsity facilitates dimensionality reduction by focusing on dominant channel components, significantly reducing feedback overhead. High feedback overhead increases system latency, computational burden, and energy consumption, while also limiting scalability in dense networks and mobile user scenarios \cite{4641946}. Accurately identifying and aligning optimal subspaces that capture channel structure while maintaining robust estimation is a complex challenge, especially in dynamic and imperfect real-world environments.

\textbf{Prior Work:} Channel estimation in sparse environments has been extensively studied through approaches such as compressive sensing (CS) and subspace-based methods. CS techniques, as explored by \cite{5454399, Baraniuk_2010}, leverage the inherent sparsity of wireless channels to reduce overhead but often suffer from high computational complexity and sensitivity to noise. These methods also require carefully designed sensing matrices and a priori knowledge of sparsity levels, limiting their practicality in dynamic, real-world environments. Subspace-based techniques \cite{7400949, 4641946}, utilize the low-rank nature of MIMO channels for efficient representation. However, these methods rely on static models or perfect channel state information, which makes them suboptimal in scenarios with imperfect or evolving channel conditions. Furthermore, both approaches often demand extensive training datasets or fail to adapt effectively to variations such as user mobility and environmental changes.

\textbf{Contribution:} Our work addresses key limitations in traditional channel estimation methods by introducing a novel framework that leverages digital twin (DT) channels as priors for subspace-based estimation. DT channels, generated through ray tracing or electromagnetic simulations, offer structured yet coarse approximations of real-world channels, capturing essential properties such as angular dispersion and power profiles \cite{10198573}. We propose a joint clustering and subspace refinement framework that dynamically adapts to changing channel conditions using user feedback. This framework operates on the Grassmannian manifold \cite{10.1145/1390156.1390204, edelman1998geometryalgorithmsorthogonalityconstraints, 7448873}, enabling iterative alignment of DT-derived subspaces with real-world characteristics, going towards the learnable digital twins \cite{jiang2024learnablewirelessdigitaltwins}. By integrating DT priors with adaptive learning mechanisms, the approach reduces computational complexity, minimizes reliance on extensive training datasets, and ensures robust and efficient channel estimation even in dynamic, sparse environments. The key contributions of this work are summarized as follows:
\begin{itemize}
    \item We propose a joint clustering and subspace refinement framework leveraging DT channels to enable low-overhead, zone-specific channel estimation.
    \item We introduce a learnable digital twin framework that integrates user feedback and iterative calibration, combining optimization on the Grassmann manifold and reinforcement learning to enhance subspace alignment.
    \item We demonstrate DT channels as effective priors, significantly reducing complexity and accelerating convergence.
\end{itemize}

\section{Signal and System Model}
\begin{figure*}[t]
    \centerline{\includegraphics[width=\textwidth]{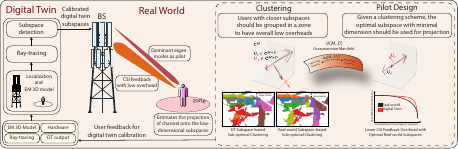}}
    \caption{The figure illustrates the proposed zone-specific subspace prediction and calibration framework for channel estimation using digital twins. The BS designs precoders for each zone, enabling UEs to estimate the projection of real-world channels onto low-dimensional DT-based subspaces. Zones are defined by user subspace similarities on the Grassmann manifold. This approach significantly reduces CSI feedback overhead by leveraging channel sparsity and DT-based subspace detection. To address DT approximation errors, subspaces are further calibrated to optimize overhead and estimation accuracy.}
    \label{fig:system}
\end{figure*}

We consider a wireless communication system with a base station (BS) equipped with a uniform planar array (UPA) of \(N = N_t N_r\) antennas, communicating with a single-antenna user equipment (UE). The wireless channel \(\mathbf{h} \in \mathbb{C}^{N}\) is modeled as a superposition of discrete propagation paths, each defined by unique angles of arrival (AoA) and departure (AoD). The channel is expressed as a linear combination of steering vectors weighted by path gains.
\begin{equation}
\mathbf{h} = \sum\nolimits_{l=1}^L \alpha_l \mathbf{a}(\theta_l, \phi_l),
\end{equation}
where \( L \) is the number of significant propagation paths, \(\alpha_l \in \mathbb{C}\) is the complex gain of the \( l \)-th path, and \(\mathbf{a}(\theta_l, \phi_l) \in \mathbb{C}^{N}\) is the array response vector associated with the azimuth angle \(\theta_l \) and elevation angle \(\phi_l\). To represent the UPA array response vector, we can use the Kronecker product as follows
\(
\mathbf{a}(\theta, \phi) =  \left(\mathbf{a}_\text{h}(\theta, \phi) \otimes \mathbf{a}_\text{v}(\phi)\right) / \sqrt{N}\),
where \( \mathbf{a}_\text{h}(\theta, \phi) \in \mathbb{C}^{N_t} \) and \(\mathbf{a}_\text{v}(\phi) \in \mathbb{C}^{N_r}\) are the horizontal and vertical steering vectors, and \( \otimes \) represents the Kronecker product.

The received signal at the UE can be written as
\begin{equation}
y = \mathbf{f}^\mathsf{H} \mathbf{h}s + n,
\end{equation}
Where \(y \in \mathbb{C}\) is the received signal, \(\mathbf{f} \in \mathbb{C}^{N}\) is the BS precoding matrix, \(s \in \mathbb{C}\) is the transmitted signal, and \(n \sim \mathcal{CN}(\mathbf{0}, \sigma^2)\) is AWGN. Sparse channel propagation in the angular domain, dominated by a few paths, allows the channel to be expressed as
\begin{equation}
\mathbf{h} = \mathbf{A}\mathbf{x},
\end{equation}
where \(\mathbf{A} \in \mathbb{C}^{N \times G}\) is an overcomplete dictionary of array response vectors, \(\mathbf{x} \in \mathbb{C}^{G \times 1}\) is a sparse representation of the channel coefficients, and \( G \) represents the discretized grid points in the angular domain.

\section{Problem Formulation}

In wireless systems, accurate channel estimation with low overhead is important for optimizing system performance. At the BS, the received signal during the channel estimation phase is modeled as
\begin{equation}
\mathbf{y} = \mathbf{F}^\mathsf{H} \mathbf{A}\mathbf{x}s + \mathbf{n},
\end{equation}
where \(\mathbf{F} \in \mathbb{C}^{N \times M}\) and \(M\) is the number of channel measurements. The task of estimating the sparse vector \(\mathbf{x}\) from the measurements \(y\) is formulated as a sparse recovery problem
\begin{equation}
\min_{\mathbf{x}} \|\mathbf{x}\|_0 \quad \text{subject to} \quad \|\mathbf{y} - \mathbf{F}^\mathsf{H} \mathbf{A}\mathbf{x}s\|_2 \leq \epsilon,
\end{equation}
where \(\|\mathbf{x}\|_0\) is the number of non-zero elements in \(\mathbf{x}\), and \(\epsilon\) is a noise tolerance threshold.

Sparse channels consist of a few dominant angular components, making it feasible to compress channel information without significant loss. While methods such as compressive sensing and autoencoders \cite{guo2019convolutionalneuralnetworkbased} have been proposed to leverage this sparsity, their implementation often incurs high computational costs. Therefore, efficient methods are needed to reduce the overhead without compromising the accuracy of channel estimation.

The performance of the reconstructed channel is evaluated using several metrics. The normalized mean squared error (NMSE) quantifies the reconstruction accuracy as
\begin{equation}
\text{NMSE} = \frac{\|\mathbf{h} - \hat{\mathbf{h}}\|_2^2}{\|\mathbf{h}\|_2^2},
\end{equation}
where \(\hat{\mathbf{h}} = \mathbf{A}\hat{\mathbf{x}}\) represents the reconstructed channel. Another metric is cosine similarity, which evaluates the alignment between the true and estimated channels
\begin{equation}
    \text{Cosine similarity} = \frac{\vert\mathbf{h}^\mathsf{H} \hat{\mathbf{h}}\vert}{\|\mathbf{h}\|_2 \|\hat{\mathbf{h}}\|_2}.
\end{equation}
Feedback overhead is also an important metric and is calculated as the total number of bits required to encode the indices of the non-zero elements and the quantized values of \(\hat{\mathbf{x}}\). This is expressed as \(B_{\text{idx}} + B_{\text{val}}\), where \(B_{\text{idx}}\) is the number of bits used to encode the indices and \(B_{\text{val}}\) represents the bits used to quantize the corresponding values.

The optimization problem for jointly minimizing the channel reconstruction loss and feedback overhead while ensuring practical feasibility can be expressed in a standard format as
\begin{subequations} \label{eq:opt_general}
\begin{align} \label{eq:opt1}
& \min_{\mathbf{F}, \hat{\mathbf{x}}, \mathcal{Q}} \hspace{.1cm} \mathcal{L} (\mathbf{h}, \hat{\mathbf{h}}) + \lambda \, \text{Overhead}(\hat{\mathbf{x}}, \mathcal{Q}) \\
& \hspace{.25cm} \text{s.t.} \hspace{0.3cm} \|\mathbf{F}\|_F^2 \leq P_{\text{BS}}, \\
& \hspace{.95cm} \|\hat{\mathbf{x}}\|_0 \leq K, \\
& \hspace{.95cm} \mathcal{Q}(\hat{\mathbf{x}}) \in \mathcal{C},
\end{align}
\end{subequations}
where \( \mathcal{L}(\mathbf{h}, \hat{\mathbf{h}}) \) is the loss function quantifying the reconstruction error between the true channel \( \mathbf{h} \) and the reconstructed channel \( \hat{\mathbf{h}} = \mathbf{A}\hat{\mathbf{x}} \). This can represent metrics like NMSE or another suitable distance measure. \( \text{Overhead}(\hat{\mathbf{x}}, \mathcal{Q}) \) captures the feedback overhead associated with the quantization \( \mathcal{Q}(\hat{\mathbf{x}}) \), including bits for indices and values. \( \|\mathbf{F}\|_F^2 \leq P_{\text{BS}} \) ensures the combining matrix adheres to the BS power constraint.
\( \|\hat{\mathbf{x}}\|_0 \leq K \) imposes sparsity on the reconstructed channel coefficients, leveraging the channel's sparse nature. \( \mathcal{Q}(\hat{\mathbf{x}}) \in \mathcal{C} \) ensures that the quantized representation \( \hat{\mathbf{x}} \) belongs to a predefined set of allowable beams, maintaining feedback feasibility.

Channel estimation overhead can be reduced by leveraging the dominant subspace of the channel matrix, representing the high-dimensional channel vector \( \mathbf{h} \in \mathbb{C}^{N_t N_r \times 1} \) with a low-dimensional subspace. The covariance matrix, computed as  
\begin{equation} 
\mathbf{R} = \frac{1}{U} \sum\nolimits_{u=1}^U \bar{\mathbf{h}}_u \bar{\mathbf{h}}_u^\mathsf{H} = \mathbf{U} \mathbf{\Sigma} \mathbf{U}^\mathsf{H},
\end{equation}  
captures the spatial structure, where \( \mathbf{U} \) and \(\mathbf{\Sigma}\) are the eigenvectors and eigenvalues. The channel vector is projected onto the \( k \)-dominant eigenvectors at the UE, reducing feedback to the coefficients \( \mathbf{z} \), and reconstructed at the BS as follows
\begin{equation} \label{eq:subspace}
\mathbf{z} = \mathbf{U}_k^\mathsf{H} \mathbf{h}, \quad \hat{\mathbf{h}} = \mathbf{U}_k \mathbf{z}.
\end{equation}  
In high-frequency bands, angular-domain sparsity allows low-rank approximations using dominant eigenvectors, minimizing reconstruction error \(\mathcal{L}(\mathbf{h}, \hat{\mathbf{h}})\) while ensuring low feedback overhead. The problem (\ref{eq:opt_general}) is reformulated as  
\begin{subequations} \label{eq:opt_k}
\begin{align}
& \min_{\mathbf{U}_k} \hspace{.3cm} \text{rank}\{\mathbf{U}_k\} \\ 
& \hspace{.2cm} \text{s.t.} \quad \mathcal{L}(\mathbf{h}, \mathbf{U}_k \mathbf{U}_k^\mathsf{H} \mathbf{h}) \leq \varepsilon, \\ 
& \hspace{.95cm} \mathbf{U}_k^\mathsf{H} \mathbf{U}_k = \mathbf{I}_k,
\end{align}
\end{subequations}  
ensuring minimal subspace rank \( k \) (\(\text{rank}\{\mathbf{U}_k\}=k\)) while maintaining reconstruction quality.

Zone-specific subspace estimation is essential as different parts of a site exhibit varying propagation characteristics, leading to subspaces with different ranks. While covariance matrices suffice for fixed zones, identifying optimal subspaces for dynamic zone partitioning requires actual channel realizations. However, with current approaches, this process can be highly costly due to the need for extensive real-world channel measurements and frequent high-overhead interactions with users to gather the required information. This underscores the importance of developing adaptive frameworks that minimize these costs while balancing overhead and performance optimization, as formulated in problem (\ref{eq:opt_k}).

\section{Proposed Digital Twin-Based Solution}

Digital twin channels provide a structured, computationally efficient means of approximating real-world wireless channels. By leveraging electromagnetic (EM) 3D models and ray-tracing techniques, DTs simulate the propagation environment, capturing dominant interactions like reflection, diffraction, and scattering. These simulations generate coarse-grained channel approximations that share key structural characteristics with real-world channels, such as spatial sparsity and multipath effects, making them invaluable for channel estimation tasks.

\subsection{Key Idea: Subspace Approximation with DT Channels}

One of the critical insights in leveraging DT channels is their ability to approximate the dominant subspaces of real-world channels. The DT covariance matrix, derived from simulated channels, captures the energy distribution across spatial dimensions, enabling the identification of principal eigenvectors. These eigenvectors span a subspace that represents the most significant directions of channel energy. The proximity of DT-based subspaces to their real-world counterparts determines the quality of channel estimation and feedback reduction.
To quantify the closeness between the subspaces of DT and real-world channels, we analyze the principal angles between these subspaces using the \textit{Kahan-Davis Sin-Theta Theorem} \cite{a6ee9c48-1e5b-385a-93c1-cdf3b873de37}. This theorem provides a bound on the misalignment of subspaces based on the spectral properties of their covariance matrices. 

\subsection{Reliability of Digital Twins Subspaces}

Let \(\mathbf{R}_{\text{DT}} \in \mathbb{C}^{N \times N}\) and \(\mathbf{R}_{\text{RW}} \in \mathbb{C}^{N \times N}\) represent the covariance matrices of the DT and real-world (RW) channels in a zone. Using eigenvalue decomposition, the \(k\)-dimensional subspaces spanned by the leading eigenvectors are denoted as \(\mathbf{U}_{\text{DT}, k}\) and \(\mathbf{U}_{\text{RW}, k}\).  
The misalignment between these subspaces is bounded by the \textit{Kahan-Davis Sin-Theta Theorem}
\begin{equation}
\sin \theta_k \leq \frac{\|\mathbf{R}_{\text{DT}} - \mathbf{R}_{\text{RW}}\|_2}{\Delta_k},
\end{equation}
where \(\Delta_k = \lambda_k(\mathbf{R}_{\text{RW}}) - \lambda_{k+1}(\mathbf{R}_{\text{RW}})\) is the spectral gap. A large \(\Delta_k\) ensures robustness, making DT subspaces reliable approximations despite \(\mathbf{R}_{\text{DT}}\) being a coarse estimate.  For small principal angles (\(\sin \theta_k \approx \theta_k\)), we have $
\theta_k \leq \|\mathbf{R}_{\text{DT}} - \mathbf{R}_{\text{RW}}\|_2/\Delta_k $ as the upperbounds. The Grassmann distance between subspaces is given by  
\begin{equation}
d_g(\mathbf{U}_{\text{DT}}, \mathbf{U}_{\text{RW}}) = \|\mathbf{\theta}\|_2^2,
\end{equation} 
where \(\mathbf{\theta} = [\theta_1, \theta_2, \dots, \theta_k]\) are the principal angles. These angles are computed as \(\theta_i = \arccos(\sigma_i)\), where \(\sigma_i\) are the singular values of \(\mathbf{U}_{\text{DT}, k}^\mathsf{H} \mathbf{U}_{\text{RW}, k}\). Smaller principal angles and Grassmann distances indicate higher subspace similarity, enhancing channel reconstruction and beamforming performance. By ensuring small Grassmann distances, DT-derived subspaces effectively approximate real-world subspaces, validating their use as priors in subspace-based estimation.

\section{Digital Twins as Prior Knowledge}

Building on the similarity between DT and real-world subspaces, DT channels serve as effective priors for channel estimation. Users with similar subspaces are grouped into zones to enable zone-specific subspace estimation, minimizing the average subspace rank required to achieve a given reconstruction loss threshold. With DT channels, the BS computes optimal low-dimensional subspaces for each zone, significantly reducing overhead, as depicted in Fig. \ref{fig:system}.

However, as DT subspaces approximate real-world channels, inaccuracies introduce errors in clustering and subspace computation. A joint optimization framework is required to address this interplay, formulated as
\begin{subequations}
\begin{align}
& \min_{\{\mathcal{C}_z\}, \{\mathbf{U}_z\}} \sum\nolimits_{z=1}^Z \text{rank}\{\mathbf{U}_z\}, \\ 
& \hspace{.55cm}\text{s.t.} \hspace{.5cm} \mathcal{L}(\mathbf{h}_u, \hat{\mathbf{h}}_u; \mathbf{U}_z) \leq \varepsilon_z, \quad \forall z, \\ 
& \hspace{1.5cm} \mathcal{C}_z \cap \mathcal{C}_{z'} = \emptyset, \quad \cup_{z=1}^Z \mathcal{C}_z = \mathcal{U}, \\ 
& \hspace{1.5cm} \|\mathbf{U}_z^\mathsf{H} \mathbf{U}_z - \mathbf{I}_{k_z}\|_F^2 \leq \epsilon, \quad \forall z, \\ 
& \hspace{1.5cm} \sum\nolimits_{z=1}^Z \sum\nolimits_{u \in \mathcal{C}_z} T_{u, z} \leq T_{\text{max}},
\end{align}
\end{subequations}
where, \(\mathcal{C}_z\) denotes the users in zone \(z\), and \(\mathbf{U}_z\) is the subspace of rank \(k_z\). The constraints enforce disjoint clustering, orthonormal subspaces, and mobility limits \(T_{u, z}\) to reduce transitions and recalculation overhead. Since DT subspaces are close to real-world ones, calibration is efficient due to the reduced search space, enabling faster convergence and accurate zone-specific channel estimation.

\subsection{Clustering on the Grassmann Manifold}

We adopt a two-step clustering framework to efficiently form subspace-aware zones. A direct one-step approach with \( k \)-means would require manual modification of its loss function to incorporate subspace distances, while \( k \)-medoids, which directly accepts distance matrices, is computationally prohibitive for large user datasets. To address this, we first apply \( k \)-means to group users into \( Z^\prime \) fine clusters (e.g., \( Z^\prime = 300 \)) based on position \cite{zhang2024zonespecificcsifeedbackmassive}. Each fine cluster's subspace is derived from its DT covariance matrix, capturing at least \( p\% \) of the total channel energy. With a significantly reduced input size, we compute a \((Z^\prime, Z^\prime)\) distance matrix using Grassmann and positional distances and apply \( k \)-medoids to merge fine clusters into larger zones (e.g., 8 zones). This hybrid approach enables zone-specific subspace estimation with minimal feedback. Further calibration is needed to align these subspaces with real-world channels, as discussed next.

\subsection{Subspace Calibration}

Subspace refinement mitigates DT approximation errors, ensuring accurate clustering and alignment of final zone subspaces with real-world channels. The goal is to optimize subspaces to capture key channel characteristics while minimizing estimation loss and feedback overhead. Building on DT-based robust frameworks \cite{10694568, 10757495, alikhani2024largewirelessmodellwm} and learnable digital twins \cite{jiang2024learnablewirelessdigitaltwins}, we propose three key strategies:  

\textbf{1. Subspace rank calibration:} After final clustering (e.g., \( k \)-medoids into eight zones), the subspace dimension \( k_z \) is adjusted to meet a performance threshold (e.g., \(-20\) dB NMSE), enhancing estimation accuracy.  

\textbf{2. Joint calibration:} Subspace tuning and clustering are refined iteratively. Fine clusters (e.g., 300 via \( k \)-means) are merged based on weighted Grassmann and positional distances, with user feedback guiding subspace updates and zone recalculations until convergence.  

\textbf{3. Subspace calibration:} Established zone subspaces are iteratively refined using user feedback to minimize reconstruction loss, ensuring robust and accurate representations for channel projection with minimal feedback overhead. In this work, we adopt this direction for DT calibration.

\textbf{Feedback mechanism:}  
In compliance with 3GPP standards \cite{3gpp.38.214}, users provide feedback on channel metrics, such as received power, to refine subspaces based on real-world channel characteristics. The loss function (e.g., NMSE or negative cosine similarity) is evaluated as a function of real-world channel power, guiding iterative subspace rotation and scaling to minimize the loss. The process continues until the loss stabilizes, indicating optimal alignment with real-world subspaces. These refined subspaces are then used to design precoders for projecting channels onto lower dimensions, achieving high performance with minimal feedback overhead.
The BS can facilitate this feedback mechanism by enabling the necessary computation at the UE.

\textit{NMSE feedback:}
The NMSE measures the residual error between the real-world channel \(\mathbf{h}_{\text{RW}}\) and the subspace-projected channel \(\mathbf{h}_{\text{SS}} = \mathbf{U}_k \mathbf{U}_k^\her \mathbf{h}_{\text{RW}}\). We have 
$
\|\mathbf{h}_{\text{RW}} - \mathbf{h}_{\text{SS}}\|^2 = \|(\mathbf{I} - \mathbf{U}_k \mathbf{U}_k^\her) \mathbf{h}_{\text{RW}}\|^2.
$
The total power of \(\mathbf{h}_{\text{RW}}\) is decomposed into the power in the dominant subspace and the residual power as 
$
\|\mathbf{h}_{\text{RW}}\|^2 = \|\mathbf{h}_{\text{SS}}\|^2 + \|(\mathbf{I} - \mathbf{U}_k \mathbf{U}_k^\her) \mathbf{h}_{\text{RW}}\|^2.
$
Substituting \(\|\mathbf{h}_{\text{SS}}\|^2 = \|\mathbf{U}_k^\her \mathbf{h}_{\text{RW}}\|^2\) and isolating NMSE, NMSE can be computed as
\(
\text{NMSE} = 1 - \|\mathbf{h}_{\text{SS}}\|_2^2 / \|\mathbf{h}_{\text{RW}}\|_2^2.
\)
To evaluate NMSE at the base station (BS), the total power \(\|\mathbf{h}_{\text{RW}}\|^2\) is fed back by the UE. The BS computes \(\|\mathbf{h}_{\text{SS}}\|^2\) locally, enabling NMSE evaluation.

\textit{Cosine similarity feedback:}
Cosine similarity quantifies the alignment between \(\mathbf{h}_{\text{RW}}\) and \(\mathbf{h}_{\text{SS}}\). We have 
$
\vert\mathbf{h}_{\text{RW}}^\her \mathbf{h}_{\text{SS}}\vert = \vert\mathbf{h}_{\text{RW}}^\her \mathbf{U}_k \mathbf{U}_k^\her \mathbf{h}_{\text{RW}}\vert = \|\mathbf{U}_k^\her \mathbf{h}_{\text{RW}}\|_2^2.
$
Substituting \(\|\mathbf{U}_k^\her \mathbf{h}_{\text{RW}}\|_2^2=\|\mathbf{U}_k \mathbf{U}_k^\her \mathbf{h}_{\text{RW}}\|_2^2\) and \(\mathbf{h}_{\text{SS}} = \mathbf{U}_k \mathbf{U}_k^\her \mathbf{h}_{\text{RW}}\), the cosine similarity can be computed as
\(
\text{cosine similarity} = \|\mathbf{h}_{\text{SS}}\|_2 / \|\mathbf{h}_{\text{RW}}\|_2.
\)

To enable efficient feedback, an augmented pilot matrix that includes the dominant subspace \(\mathbf{U}_k\) and its orthogonal complement could be used.

\subsection{RL-Based Subspace Calibration} \label{subsec:drl}

Aligning digital twin subspaces with their real-world counterparts is challenging due to the high-dimensional nature of wireless channels and the complex relationships between DT and real-world representations. Wireless channels exhibit angular-domain sparsity, with dominant multipath components confined to a small subset of \textbf{discrete Fourier transform (DFT)} codebook vectors (beams) \cite{7160780} within each zone. Let \(\mathbf{F} \in \mathbb{C}^{N \times N}\) denote the DFT matrix, and let \(\mathbf{x} \in \mathbb{C}^N\) be the sparse angular-domain representation of the channel satisfying \(\mathbf{h} = \mathbf{F} \mathbf{x}\). A \textbf{majority voting} mechanism is employed to identify the most frequently occurring DFT beams across DT channels within a zone, ranking them in order of importance. Since these dominant beams are directly linked to the zone’s subspace orientation, calibrating DT-based subspaces involves aligning these beams with their real-world counterparts. However, selecting the optimal \( k_z \) dominant beams from an \( N \)-dimensional DFT codebook requires evaluating \( \binom{N}{k_z} \) possible configurations, which becomes computationally prohibitive for large \( N \) or dense deployments. Furthermore, deep learning-based approaches necessitate extensive labeled data, which is often infeasible to obtain in practical settings.  

To address this, we formulate the problem as a sequential decision-making task and employ a \textbf{deep reinforcement learning (DRL)} framework for iterative subspace refinement. The DRL agent learns an optimal alignment policy by interacting with users in a zone and receiving real-time power measurement feedback, which are mapped to the average cosine similarity within each zone. The optimization process is modeled as a Markov decision process (MDP), where the state \(\mathbf{s}_t \in \{0,1\}^{N-10} \times \mathbb{R}^{10}\) consists of a binary mask representing active beams along with a 10-step history of subspace alignment metrics. At each step, the agent replaces a selected beam \( b_i \in \mathcal{B}_t \) with an unused beam \( b_j \) following an initialization-dependent replacement strategy as follows 
\begin{equation}
b_i =
\begin{cases} 
\underset{b \in \mathcal{B}_t}{\arg\min}\ \mathbb{E}\left[\|\mathbf{F}_b^H \mathbf{h}_{\text{DT}}\|^2\right], & \text{DT-based}, \\
\text{Uniform}(\mathcal{B}_t), & \text{Random}.
\end{cases}
\end{equation}
The agent is trained using a reward function that encourages subspace alignment improvements while penalizing performance degradation given by
\begin{equation}
r_t = \text{clip}\left(\frac{\mathcal{S}_{t+1} - \mathcal{S}_t}{|\mathcal{S}_0|}, -1, 1\right) - 0.5 \cdot \mathbb{I}(\mathcal{S}_{t+1} < \mathcal{S}_0),
\end{equation}
where \( \mathcal{S}_t \) quantifies the average cosine similarity within the zone. The training process is based on a clipped Double Deep Q-Network (DDQN) architecture with twin Q-networks, \( Q_{\text{online}} \) and \( Q_{\text{target}} \), updated as 
\begin{equation}
    \begin{aligned}
    Q_{\text{target}}(s_t,a_t) &\leftarrow r_t + \gamma 
    \max_{a'} Q_{\text{target}} \Big(s_{t+1}, \\
    &\hspace{.56cm} \underset{a}{\arg\max} \, Q_{\text{online}}(s_{t+1},a) \Big).
    \end{aligned}
\end{equation}    
To enhance stability, gradient clipping is applied with \( \|\nabla Q\|_2 \leq 1 \), and an adaptive exploration rate follows an exponential decay schedule: \( \epsilon \leftarrow \max(0.1, 0.9995^t) \).  

To ensure scalability, a multi-agent reinforcement learning framework is adopted, where each zone operates an independent DRL agent. This decentralized approach enables parallel learning and adaptation, allowing policies to be tailored to the unique propagation characteristics of each zone. Given a DFT codebook of dimension \( N \), the computational complexity of the proposed calibration framework scales as \( \mathcal{O}(TZN^2) \) across \( T \) training episodes and \( Z \) zones.

\section{Simulation}

We consider a 128-dimensional UPA at the BS, serving single-antenna users in the mmWave band. Real-world channels are modeled using the Indianapolis scenario of the DeepMIMO dataset \cite{alkhateeb2019deepmimogenericdeeplearning}, with a maximum of \(3\) reflections. To simulate digital twins, we introduce perturbations by randomly shifting buildings \(4\) meters and performing ray tracing with Wireless InSite \cite{Remcom}. In the DT scenario, users experience at most \(1\) propagation path, while in the real world, this increases to \(25\). These perturbations and DT's lower fidelity introduce inaccuracies, particularly in the AoD, causing misalignment between DT and real-world beams in the DFT codebook. The SNR is set to \(10\) dB.

\subsection{Subspace Detection for Channel Estimation} 
\begin{figure}[t]
    \centerline{\includegraphics[width=\columnwidth]{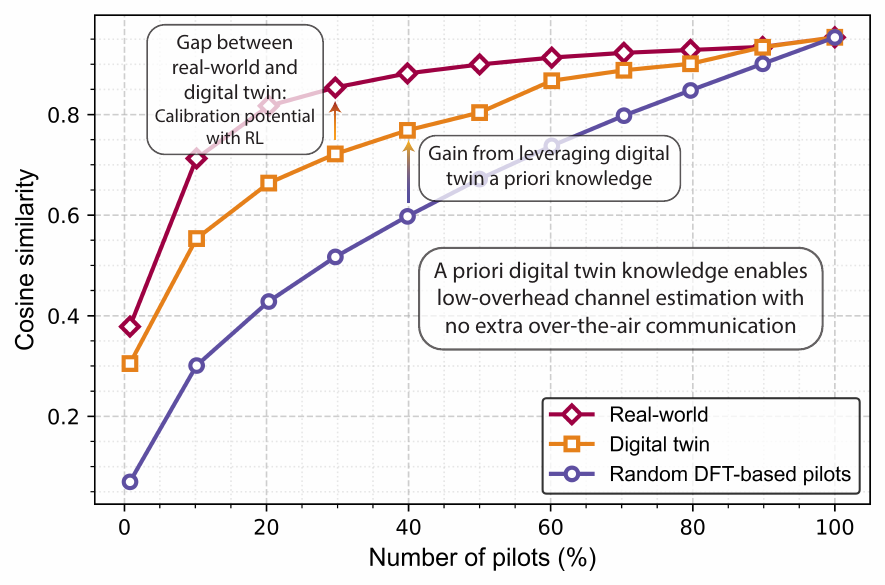}}
    \caption{Cosine similarity of channel estimation vs. average pilot usage, determined by subspace rank and power coverage per zone. Subspace ranks vary across zones, and the figure shows their averages.}
    \label{fig:perf_vs_pilots}
\end{figure}

The simulation evaluates the proposed framework for channel estimation by comparing different pilot design strategies. The process begins with \( k \)-means clustering, segmenting the site into \( Z' = 80 \) fine clusters, which are then merged into \( Z = 12 \) final zones using \( k \)-medoids, leveraging both Grassmannian and spatial distances. For each zone, dominant DFT beams are identified via majority voting on DT channels and subsequently used as pilots for low-overhead CSI feedback at the UE. Fig.~\ref{fig:perf_vs_pilots} illustrates the cosine similarity across different pilot overhead levels for various zones.  

Three pilot selection approaches are considered: (i) \textbf{Real-world dominant beams}, serving as an optimal but impractical benchmark due to the BS’s lack of precise subspace knowledge; (ii) \textbf{DT-based dominant beams}, which utilize prior DT knowledge to estimate dominant beams and approximate subspaces, significantly reducing pilot overhead; and (iii) \textbf{random DFT beams}, acting as a baseline \cite{6181796, 5454399, 10878425, haghshenas2023efficientloschannelestimation}, demonstrating the inefficiency of \textit{uninformed pilot selection}. 

Cosine similarity provides a scale-invariant measure of subspace alignment, making it a more suitable performance metric than NMSE, as it eliminates the need for magnitude calibration. This motivates its use in this work. In the low-similarity regime, achieving a similarity of \(0.8\) requires fewer than \(20\%\) of pilots when DT-selected beams are perfectly accurate. However, due to DT approximation errors, this requirement increases to \(50\%\) of the \(128\) pilots, while random DFT beams demand \(70\%\). In the high-similarity regime (\(0.9\) target) at \(10\) dB SNR, DT-based selection requires \(80\%\) of pilots compared to \(50\%\) for real-world-based selection, while random DFT beams remain inefficient, requiring \(90\%\). The significant performance gains observed across some consecutive steps stem from the fact that DFT beams are not equally important within each zone—some beams are more frequently selected as the best beam and thus contribute more to the overall signal propagation characteristics. These findings highlight the advantage of prioritizing more contributive beams, leading to more efficient pilot allocation and improved calibration.

\subsection{RL-Based Subspace Calibration}

\begin{figure}[t]
    \centerline{\includegraphics[width=\columnwidth]{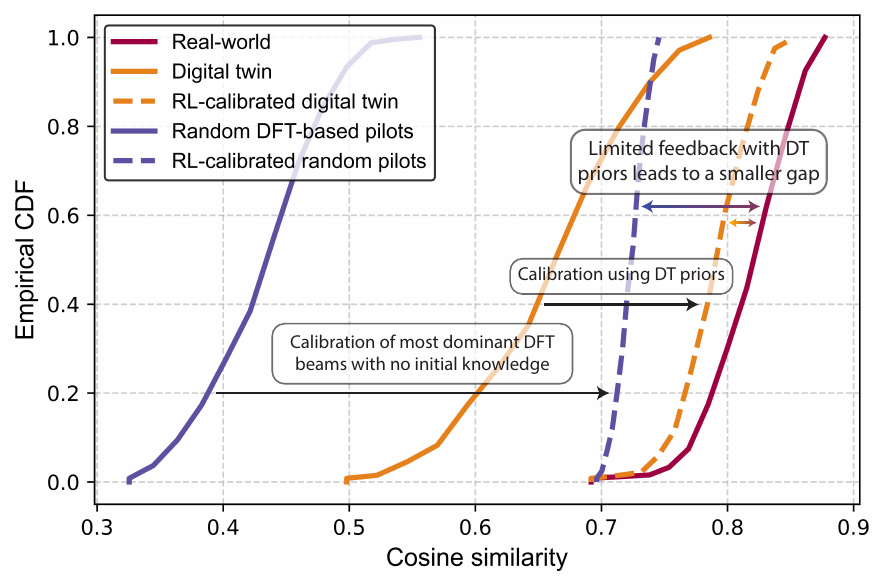}}
    \caption{Reinforcement learning bridges the performance gap between zone-specific DT and RW subspaces by leveraging digital twin knowledge as a prior and integrating real-time reward feedback, advancing learnable digital twin models.}
    \label{fig:cdf}
\end{figure}  

To refine DT-based per-zone dominant beams, we employ the DRL-based calibration algorithm introduced in Section~\ref{subsec:drl}. Given the practical constraint of limited user feedback, we restrict the number of training episodes to \( 300 \) and evaluate two key aspects: (i) the effectiveness of the DT-based beam calibration using real-time user feedback and (ii) the advantage of DT-based initialization over the random initialization method used in \cite{6181796, 5454399, 10878425, haghshenas2023efficientloschannelestimation}.  

In this process, we leverage DT knowledge as prior information for DRL calibration by incorporating the order of the most contributive beams within each zone. This allows the DRL model to start with a structured initialization, prioritizing beams that are more influential in the DT approximation. The evaluation is performed with a fixed pilot allocation of \( 20\% \) of the total \( 128 \) pilots, assessing performance through the cumulative distribution function (CDF), as shown in Fig.~\ref{fig:cdf}. The performance variability across trials is attributed to channel estimation noise. The DRL-based calibration of DT-derived dominant beams achieves significant improvements in convergence within a limited number of episodes. This result underscores the potential of reinforcement learning in systematically bridging the gap between digital twins and real-world channel subspaces, enabling efficient and adaptive calibration over time.

\section{Conclusion}

This paper proposes a framework for zone-specific channel estimation using digital twins as priors, leveraging mmWave channel sparsity. A two-step clustering process with reinforcement learning refines DT-based subspaces to align with real-world channels using user feedback. The approach reduces feedback overhead and enhances estimation accuracy, showcasing DTs as effective starting points for subspace-based estimation and advancing adaptive wireless systems.

\bibliographystyle{ieeetr} 
    
\end{document}